\documentclass[10pt]{article}

\def\Title#1{\begin{center} {\Large #1 } \end{center}}
\def\Author#1{\begin{center}{ \sc #1} \end{center}}
\def\Address#1{\begin{center}{ \it #1} \end{center}}

\newcommand\pubblock{\rightline{\begin{tabular}{l} Proceedings of the Second Annual LHCP\\ \pubnumber\\
         \pubdate  \end{tabular}}}

\newenvironment{Abstract}{\begin{quotation} \begin{center} 
             \large ABSTRACT \end{center}\bigskip 
      \begin{center}\begin{large}}{\end{large}\end{center} \end{quotation}}

\newenvironment{Presented}{\begin{quotation} \begin{center} 
             PRESENTED AT\end{center}\bigskip 
      \begin{center}\begin{large}}{\end{large}\end{center} \end{quotation}}

\usepackage{color}
\usepackage{times}
\usepackage{amsmath,amssymb}
\usepackage{caption}
\usepackage{subcaption}

\usepackage{amssymb}
\usepackage{amsmath}
\usepackage{amssymb}
\usepackage{graphicx}
\usepackage{float}
\usepackage{braket}
\usepackage{url}
\usepackage{cite}
\usepackage{subcaption}

\newcommand\pubnumber{ CMS CR-2014/190 }

\newcommand\pubdate{\today}

\def\affiliation{
On behalf of the CMS Collaboration, \\
Department of Physics \\
Brown University, Providence, RI 02912, U.S.A }

\def\Lxy{L_{xy}}

\begin{document}

\large
\begin{titlepage}
\pubblock

\vfill
\Title{  Measurement of the Top Quark Mass With 2012 CMS Data }
\vfill

\Author{ Richard Nally  }
\Address{\affiliation}
\vfill
\begin{Abstract}

The mass of the top quark was an active topic of research at CMS using 2011 data, and remains so as the 2012 data analysis campaign proceeds. Here we discuss some of the earliest results on the top mass using 2012 $\sqrt{s} = 8$ TeV CMS data, including measurements of the top mass from semileptonic $t\bar{t}$ decays and the lifetime of the $B$-hadron, as well as a measurement of the top-antitop mass difference. 

\end{Abstract}
\vfill

\begin{Presented}
The Second Annual Conference\\
 on Large Hadron Collider Physics \\
Columbia University, New York, U.S.A \\ 
June 2-7, 2014
\end{Presented}
\vfill
\end{titlepage}
\def\thefootnote{\fnsymbol{footnote}}
\setcounter{footnote}{0}
%

\normalsize 


\section{Introduction}

The mass of the top quark is one of the nineteen empirical parameters of the Standard Model (SM), and is an active topic of research at CMS. The top mass is an input to global electroweak fits, and is linked by radiative corrections to the mass of the Higgs and W bosons, making its measurement crucial for understanding both SM and beyond SM physics. Combined 2011 CMS measurements of the top quark mass found $m_t$ =173.36 $\pm$ 0.38 (stat.) $\pm$ 0.91 (syst.) GeV \cite{2011Combination}. Here, we discuss several early measurements of or relating to the top mass using data collected in 2012 by the CMS detector, described in \cite{Detector}. 

\section{Measuring the Top Mass from $B$-Hadron Decays}

In CKM-favored $t\to Wb$ decays, $B$ hadrons are formed by the $b$-quark. In such events, it can be shown through an analysis of event kinematics that the decay length $L_{xy}$ of this $B$ hadron is directly proportional to $m_t$. In particular, $\Lxy = \gamma_B\beta_B\tau_B \propto 0.4\frac{m_t}{m_b}\beta_B\tau_B$ \cite{Bhadron}. It was hence proposed to measure the top mass from distributions of $\Lxy$ in $B$-hadron decays as a supplement to more traditional measurement methods. Importantly, such an analysis uses information from CMS's tracker, instead of its calorimeter, and hence avoids systematics from the jet energy scale (JES), which dominate other measurements. 

\begin{figure}[htb]
\begin{subfigure}{.33\textwidth}
\begin{centering}
\includegraphics[scale=.6]{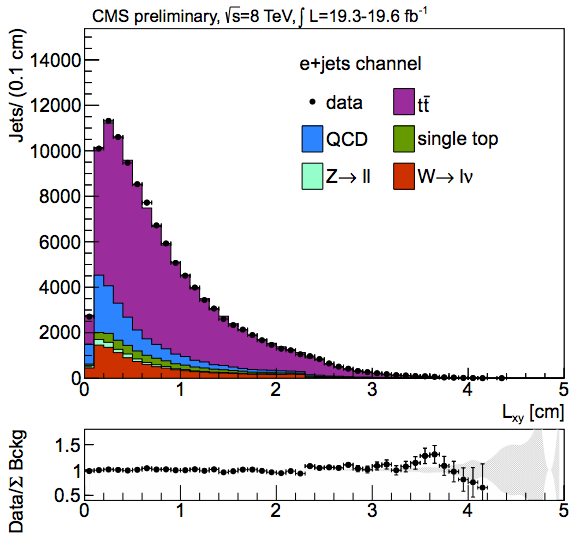}
\caption{Distribution of $\Lxy$ in the $e$+jets channel.}
\end{centering}
\end{subfigure} 
~
\begin{subfigure}{.33\textwidth}
\begin{centering}
\includegraphics[scale=.6]{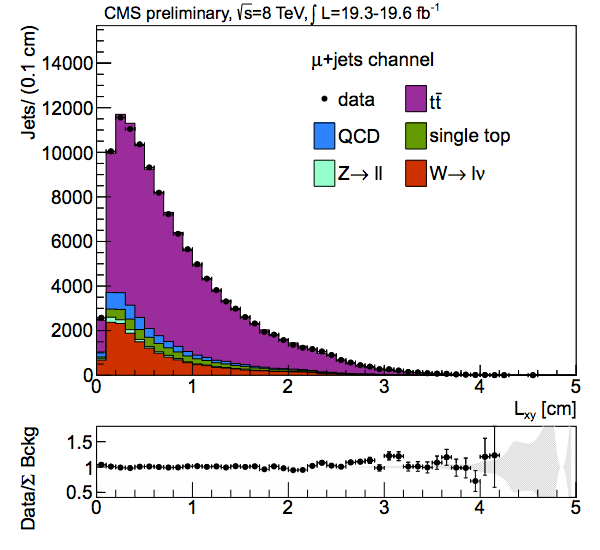}
\caption{Distribution of $\Lxy$ in the $\mu$+jets channel.}
\end{centering}
\end{subfigure}
~
\begin{subfigure}{.33\textwidth}
\begin{centering}
\includegraphics[scale=.6]{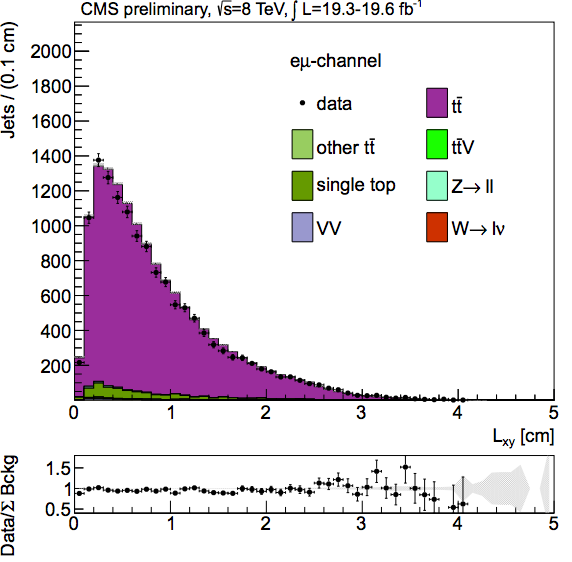}
\caption{Distribution of $\Lxy$ in the $e\mu$+jets channel.}
\end{centering}
\end{subfigure}
\caption{Distributions of $\Lxy$ in the three channels used in the analysis.}
\label{fig:LxyDistro}
\end{figure}

 Two categories of events are examined: events with one charged lepton and at least four jets, and events with exactly one electron and one muon, with at least two jets; this gives a total of three channels (singe electron, single muon, electron and muon). In each event that passes cuts, the largest $\Lxy$ associated with a secondary vertex is chosen. Distributions of $\Lxy$ are generated separately for each of the three channels. These distributions are shown in Figure \ref{fig:LxyDistro}.

 From event kinematics, we see that the mean $\widehat{L_{xy}}$ of each of these distributions has a linear dependency on $m_t$; the exact linear coefficient, however, is not given explicitly. Instead, the coefficients are found by calibrating the measurement independently in each channel. Monte Carlo (MC) distributions of $\Lxy$ are generated at several fixed values of the top mass; at each of these mass points, $\widehat{L_{xy}}$ is found for each of the three channels. Lines are then fit to these three distributions of $\widehat{L_{xy}}$, providing the needed linear coefficients. With the linear dependencies calibrated, the top mass itself is extracted from the means of the distributions. The final measured value of the top mass is the best linear unbiased estimator of the combination of the results of the three channels, and is found to be $m_t$ = 173.5 $\pm$ 1.5 (stat.) $\pm$ 1.3 (syst.) $\pm$ 2.6 $\left(p_T^t\right)$ GeV. The last uncertainty is derived from considering the distributions of top quark $p_T$ in simulation, and dominates the result.

\section{Measuring the Top Mass from Semileptonic $t\bar{t}$ Decays}
\begin{figure}[htp!]
\begin{subfigure}{.45\textwidth}
\begin{centering}
\includegraphics[scale=.57]{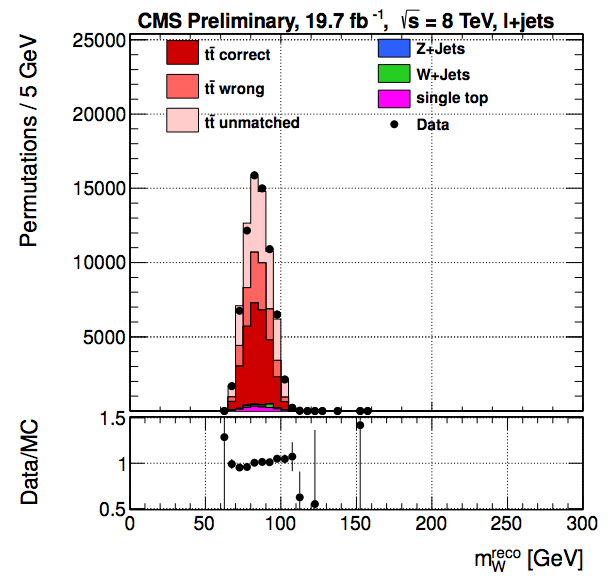}
\caption{Distribution of $M_W^{reco}$.}
\end{centering}
\end{subfigure}
~
\begin{subfigure}{.45\textwidth}
\begin{centering}
 \includegraphics[scale=.57]{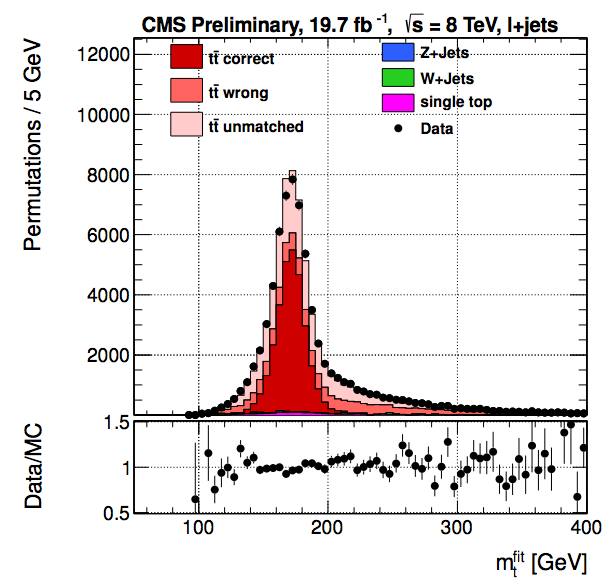}
 \caption{Distribution of $M_t^{fit}$.}
 \end{centering}
 \end{subfigure}
 \caption{Distributions of $m_W^{reco}$ and $m_t^{fit}$ in data and simulation used in the ideogram measurement.}
 \label{fig:SemiLDistros}
\end{figure}

A standard top mass measurement technique, known as the ideogram method, analyzes semileptonic $t\bar{t}$ decays, i.e. events with $t\bar{t}\to bWbW \to  b\ell\nu bq\bar{q}$. Events with exactly one charged lepton and at least four jets, with exactly two $b$-tagged jets among the four leading jets \cite{ideogram}. The mass $m_W^{reco}$ of the $W$ boson from the leptonic decay can be reconstructed directly from the measured energy and momenta of the two leading non-$b$-tagged jets. Meanwhile, a top mass estimator $m_t^{fit}$ can be obtained from a kinematic fit to the leptonic decay products. Distributions of $m_W^{reco}$ and $m_t^{fit}$ are shown in Figure \ref{fig:SemiLDistros}. 

\begin{figure}[htp!]
\begin{centering}
\includegraphics[scale=.3]{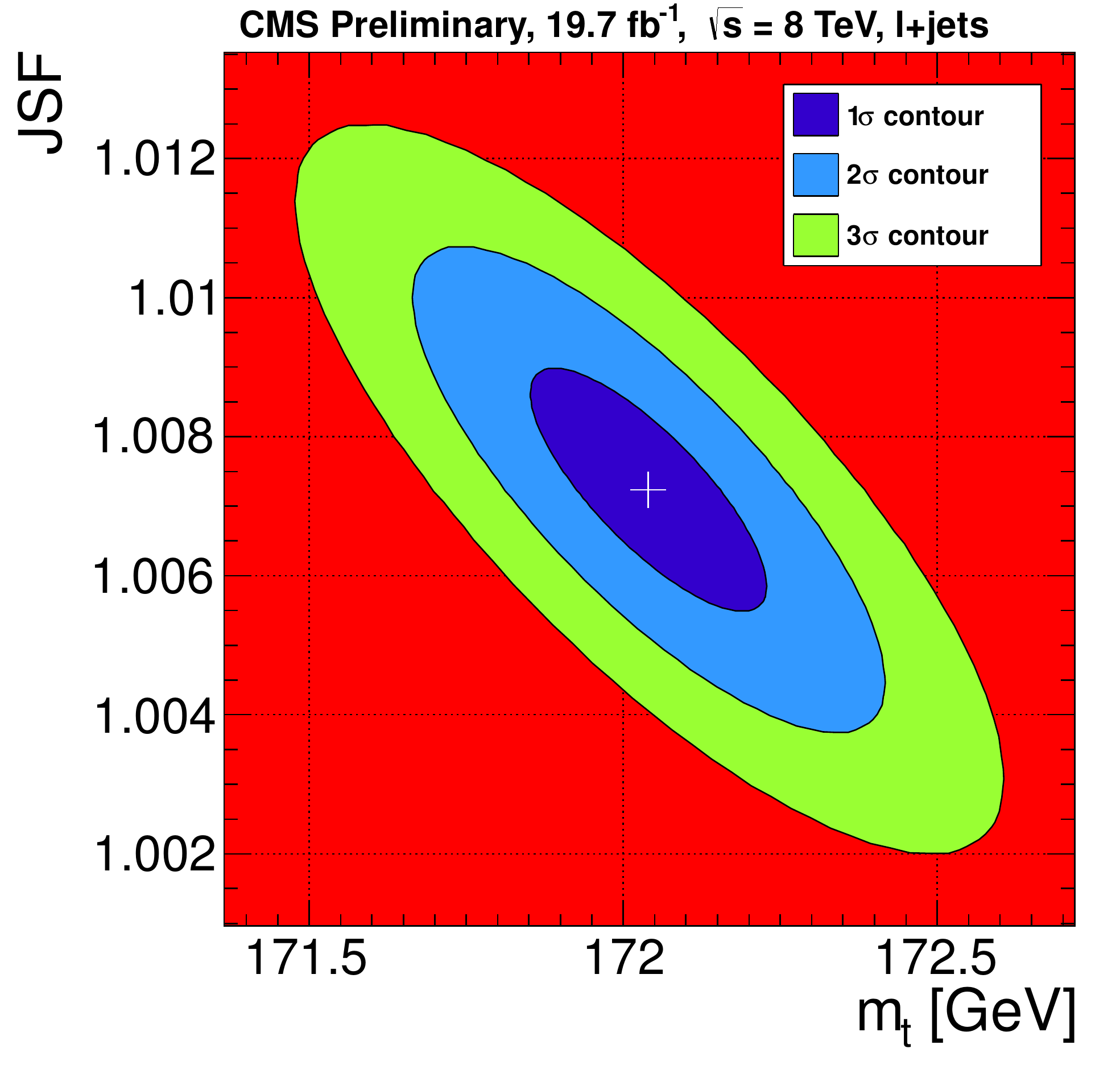}
\caption{Two-dimensional likelihood contours in $m_t$-JES space.}
\label{fig:SemiL}
\end{centering}
\end{figure}

The ideogram method is a two-dimensional likelihood fit that simultaneously measures the top mass $m_t$ and the jet scale factor (JSF), a scale factor applied to all simulation jet energies to account for the JES. In an ideogram measurement, a kinematic fit to each event yields a likelihood curve as a function of $m_t$ and the JES; the measured $m_t$ and JES are the global maxima of the product of all single-event likelihood functions. The measured likelihood contours are shown in Figure \ref{fig:SemiL}. 

The measurement finds  $m_t$ = 172.04 $\pm$ 0.19 (stat + JSF) $\pm$ 0.75 (syst.)  GeV, and JSF = 1.007 $\pm$ 0.002 (stat.) $\pm$ 0.012 (syst.). This is the first single top mass measurement with sub-GeV uncertainty, and hence marks an important achievement.

\section{Measuring the Top-Antitop Mass Difference}

\begin{figure}[H]
\begin{subfigure}{.45\textwidth}
\begin{centering}
\includegraphics[scale=.75]{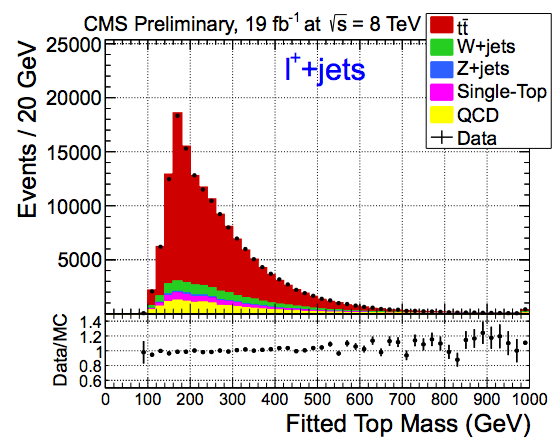}
\caption{Reconstructed top masses in events with a positive lepton.}
\end{centering}
\end{subfigure}
~
\begin{subfigure}{.45\textwidth}
\begin{centering}
\includegraphics[scale=.75]{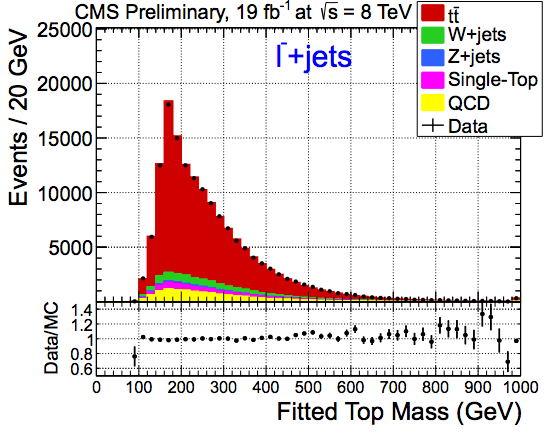}
\caption{Reconstructed top masses in events with a negative lepton.}
\end{centering}
\end{subfigure}
\caption{Distributions of reconstructed top masses in semileptonic $t\bar{t}$ events segregated by lepton sign.}
\label{fig:CPT}
\end{figure}

The ideogram method described above was also used to study the mass difference $\Delta m_t$  between the top and antitop quarks. In the ideogram method's event selections, it is easy to distinguish between events where the top and antitop quarks decay leptonically; for instance, if a top quark decays leptonically, we will have a positively charged lepton. Events passing cuts in an ideogram analysis were separated by lepton charge, yielding the reconstructed mass distributions shown in Figure \ref{fig:CPT}. An ideogram measurement was performed with the events in each of these two categories, and the results were subtracted to find the mass difference. The measurement found $\Delta m_t$ = -272 $\pm$ 196 (stat.) $\pm$ 122 (syst.)  MeV, consistent with the Standard Model prediction. 

\section{Conclusions}

We have presented three top mass results using 2012 CMS data. From these results alone, we already see single measurements of the top mass that are competitive with world combinations, as well as innovative tests of both the top mass and tests of crucial SM predictions.



\begin{thebibliography}{99}





\bibitem{2011Combination}
  CMS Collaboration,
  ``Combination of CMS results on the mass of the top quark using up to 5.0 fb$^{-1}$ of data",
  CMS PAS TOP-11-018
  
  \bibitem{Detector}
  CMS Collaboration,
  ``The CMS experiment at the CERN LHC",
  JINST 03 (2008) S08004

\bibitem{Bhadron}
   CMS Collaboration,
   ``Measurement of the top quark mass using the B-hadron lifetime technique",
   CMS PAS TOP-12-030
   
\bibitem{ideogram}
  CMS Collaboration,
  ``Measurement of the top-quark mass in $t\bar{t}$ events with lepton+jets final states in pp collisions at $\sqrt{s}$=8 TeV",
  CMS PAS TOP-14-001
  
\bibitem{CPT}
  CMS Collaboration,
  ``Measurement of the top-antitop mass difference in pp collisions at $\sqrt{s}$ = 8 TeV",
  CMS PAS TOP-14-031
   



\end{thebibliography}
\end{document}